Original Paper

# eHealth Intervention to Improve Health Habits in the Adolescent Population: Mixed Methods Study

Carmen Benavides[1*], BSc, MSc, PhD; José Alberto Benítez-Andrades[1*], BSc, MSc, PhD; Pilar Marqués-Sánchez[2*], BSc, MSc, PhD; Natalia Arias[2*], BSc, MSc, PhD

[1]SALBIS Research Group, Department of Electric, Systems and Automatics Engineering, University of León, León, Spain
[2]SALBIS Research Group, Faculty of Health Sciences, University of León, Ponferrada, Spain
[*]all authors contributed equally

**Corresponding Author:**
José Alberto Benítez-Andrades, BSc, MSc, PhD
SALBIS Research Group
Department of Electric, Systems and Automatics Engineering
University of León
Campus de Vegazana s/n
León
Spain
Phone: 34 987293628
Email: jbena@unileon.es

## Abstract

**Background:** Technology has provided a new way of life for the adolescent population. Indeed, strategies aimed at improving health-related behaviors through digital platforms can offer promising results. However, since it has been shown that peers are capable of modifying behaviors related to food and physical exercise, it is important to study whether digital interventions based on peer influence are capable of improving the weight status of adolescents.

**Objective:** The purpose of this study was to assess the effectiveness of an eHealth app in an adolescent population in terms of improvements in their age- and sex-adjusted BMI percentiles. Other goals of the study were to examine the social relationships of adolescents pre- and postintervention, and to identify the group leaders and study their profiles, eating and physical activity habits, and use of the web app.

**Methods:** The BMI percentiles were calculated in accordance with the reference guidelines of the World Health Organization. Participants' diets and levels of physical activity were assessed using the Mediterranean Diet Quality Index (KIDMED) questionnaire and the Physical Activity Questionnaire for Adolescents (PAQ-A), respectively. The variables related to social networks were analyzed using the social network analysis (SNA) methodology. In this respect, peer relationships that were considered reciprocal friendships were used to compute the "degree" measure, which was used as an indicative parameter of centrality.

**Results:** The sample population comprised 210 individuals in the intervention group (IG) and 91 individuals in the control group (CG). A participation rate of 60.1% (301/501) was obtained. After checking for homogeneity between the IG and the CG, it was found that adolescents in the IG at BMI percentiles both below and above the 50th percentile (P50) modified their BMI to approach this reference value (with a significance of $P<.001$ among individuals with an initial BMI below the P50 and $P=.04$ for those with an initial BMI above the P50). The diet was also improved in the IG compared with the CG ($P<.001$). After verifying that the social network had increased postintervention, it was seen that the group leaders (according to the degree SNA measure) were also leaders in physical activity performed ($P=.002$) and use of the app.

**Conclusions:** The eHealth app was able to modify behaviors related to P50 compliance and exert a positive influence in relation to diet and physical exercise. Digital interventions in the adolescent population, based on the improvement in behaviors related to healthy habits and optimizing the social network, can offer promising results that help in the fight against obesity.

*(JMIR Mhealth Uhealth 2021;9(2):e20217)*   doi: 10.2196/20217

## Introduction

The current trend toward rising BMIs among children and adolescents reflects a standstill in high-income countries and an increase in cases of overweight and obesity in lower-middle-income countries [1]. It is estimated that by 2025, 268 million children worldwide between the ages of 5 and 17 years could be overweight, including 91 million children who meet the criteria for obesity [2]. The World Health Organization (WHO) recognizes that social changes have modified factors related to diet and physical exercise, resulting in an energy imbalance that leads to overweight and obesity [3]. In fact, among all the determining factors that can influence diet and physical activity in the adolescent population, peers can encourage healthy or unhealthy behaviors [4].

Indeed, diet and physical exercise play a role in the conditions of overweight and obesity. In relation to diet, good adherence to the Mediterranean diet has many health benefits [5]. In fact, it is recommended that both Mediterranean and non–Mediterranean countries adhere to the principles of this type of diet to combat obesity and other chronic diseases [6,7]. The problem we currently face is that children and adolescents in European Mediterranean countries tend to neglect the Mediterranean diet pattern because of sociodemographic and lifestyle changes [8], among other causes. In Spain, for example, recent studies have shown that 64.3% of children between ages 6 and 17 years have low or medium adherence to the Mediterranean diet [9].

The same applies to physical activity. Although the benefits of exercise are widely recognized, 55.4% of Spanish children and adolescents do not comply with international recommendations on physical activity [10]. On the other hand, it has been recognized that putting an end to the obesity epidemic requires consideration of the environmental factors that play a role in the problem rather than focusing strictly on the individual as being solely responsible [11]. In fact, the broader the attack strategy, the better the results are expected to be. For example, interventions applied at the school level and focused on food and exercise in combination are considered promising in the fight against obesity [12]. Recent trends show an emerging number of technology-based, adolescent-focused interventions to improve health outcomes for multiple related behaviors—in particular, interventions through web-based platforms focused on health education that propose the fulfilment of certain proposed objectives and invite self-management, succeed in fostering parental involvement, and improve diet and physical activity behaviors [13].

In this context of applying technology to improve the diets and physical activity levels of adolescents, the social environment can be used to facilitate mechanisms of influence and social support and/or the provision of resources between individuals. To achieve this relational perspective, the social network analysis (SNA) paradigm has been applied to analyze the social environment from a structural point of view [14]. There are a number of significant studies that apply the SNA methodology to interventions in obesity and its associated factors. For example, health behavior change in favor of more physical activity could be promoted if we adapt a specific intervention to the characteristics of young people (according to their gender) and their social network [15]. Another option is to identify and work with the "leaders" among peer groups to influence their peers, offering them training so that they can propagate the desired change in behavior through their social network [16]. Several studies are currently being conducted that are trying to shed light on the ways in which interventions based on social networks can be implemented to increase their effectiveness, as it is a field of research with many avenues to explore [17]. Although it has been shown that the mechanism of influence exists by which a health behavior can spread among the individuals in a network, it is not clear how that mechanism works (eg, contagion, acceptance of group norms, imitation, etc) nor how the intervention should be designed to facilitate it [18].

In an attempt to optimize the adolescent social network, while focusing our efforts on improving healthy habits in the adolescent population, we created an eHealth web app with a responsive design called "SanoYFeliz" (ie, Healthy and Happy), which is accessible from anywhere with an internet connection. SanoYFeliz is only available in Spanish, although the English version is being developed for use in a possible international project in the future. This eHealth app, which was developed in the school environment and offers the possibility of parental participation, was based on support from peer networks and provides a system of virtual rewards for the achievement of objectives. As such, SanoYFeliz could be an enormously useful tool in the improvement of behaviors related to eating and physical activity habits. The proposed objectives of this study were (1) to assess the effectiveness of the eHealth app in the adolescent population in terms of age- and gender-adjusted BMI percentiles; (2) to study the social relationships of the adolescents pre- and postintervention; and (3) to identify the group leaders and study their profiles, eating and physical activity habits, and use of the web app.

## Methods

### Study Design

The study employed a pre-post experimental design using the technique of intentional or convenience sampling and ran for 14 weeks between October 2019 and January 2020. This study is part of the "Acquisition of healthy routines in the adolescent population with a tendency to obesity, through an automated coaching platform based on social networks and Semantic Web" project, funded by the Junta de Castilla y León (The Castilla y Leon Regional Council) in Spain. Before this intervention was conducted, a pilot study was carried out with the participation of 95 adolescents belonging to a single school in order to detect possible defects and resolve them.

### Permissions

Because this research was conducted in a population of minors, informed parental/guardian consent was requested prior to adolescent participation. Adolescent consent was implicit if they agreed to participate in the intervention; no adolescent was forced to participate. Given the educational context in which this study was developed, permission was obtained from the Department of Education of the Junta de Castilla y León. It was also approved by the Ethics Committee of the University of León (ETICA-ULE-028-2018).

## Participants

The population under study was adolescents in their first and second year of compulsory secondary education from three educational centers in the province of León, Spain. After receiving authorization from the directors of the educational centers, a total population of 340 students was assigned to the intervention group (IG; 168 students in their first year and 172 students in their second year) and 171 students to the control group (CG; 90 students in their first year and 81 students in their second year). In total, 124 students in the IG and 80 students in the CG were excluded from the study for not having the informed consent form signed by their parents. Therefore, the IG comprised 216 adolescents (119 in their first year and 97 in their second year) and the CG comprised 91 adolescents (48 in their first year and 43 in their second year). However, the final sample size of the IG was 210 students, as 6 individuals could not participate because of school absence for medical reasons. The criteria for selecting the CG and IG were based on the need for the samples not to be contaminated. As there were two schools in León and one in Ponferrada, it was decided that one of the schools in León would be the CG. In this way we ensured that there was no communication between students in the same school.

## eHealth App Development

The eHealth app, SanoYFeliz, was available to all of the adolescents participating in the study. Students were introduced to the app at the beginning of the intervention and were provided with several video tutorials explaining the different functionalities of the app. They were always supervised by the eHealth app administrators and researchers in this project, so that there was no conflict with or abuse toward any of their schoolmates. The adolescents were reminded of their total freedom to use the app without further interference. However, it is important to emphasize that the physical education teacher motivated them to use the app on a weekly basis. Also, since parents are an important source of influence for adolescents [19], it was suggested that participants involve their parents in some of the activities they performed on the app. Although more details about the app can be found in the paper presented at HEALTHINF 2020 [20], some of its functionalities include the ability to create a personal profile and network of contacts online, communicate with friends, and create healthy events and attend them. The app also provides personalized healthy eating and physical exercise tips, and rewards users virtually through "bienStars" ("healthyStars") to motivate their participation.

This eHealth app was designed using a variety of behavior change techniques, such as social support (through comments and "likes" that users can send to others), offering information and recommendations (through short messages), or gambling with virtual rewards (through "healthyStars" points) [21].

## Functionalities of the App Accessible by the CG Versus the IG

The 91 participants in the CG were part of the 14-week intervention, but they only had access to the public part of the eHealth app (ie, the part of the app that could be accessed from any device by any anonymous user) (Figure 1). The CG could not access the social part of the app (ie, the part where they could request contact or use the chat), create events, access the nutrition or physical activity tips, or receive virtual rewards. On the other hand, the IG was able to create a username and password and use all of the aforementioned features. Users of the app who have access to the restricted area (ie, the IG) can make use of all of the functionalities of the app, including the following:

- access to the social network: add friends, comment on different walls, give likes to publications, create events, and get points in the reward system (healthyStars; Figure 2); and
- personalized notifications: the app sends personalized notifications and advice about nutrition and physical activity. This is accomplished by using push notifications, which are available on smartphones, tablets, and web browsers, as well as sending emails.

Meanwhile, visitors to the website (ie, the CG) can only view the front page and project information and access the blog that contains articles on nutrition and physical activity of approximately 1000 words.

It should be noted that the students in the IG had access to the eHealth app on a daily basis, including all of the functionalities, while students in the CG could only access the public part mentioned above (no access to the social networks, personalized notifications, reward systems, etc).

**Figure 1.** Screenshot of SanoYFeliz showing the public interface (left and right panels) and the account menu of a user (center panel).

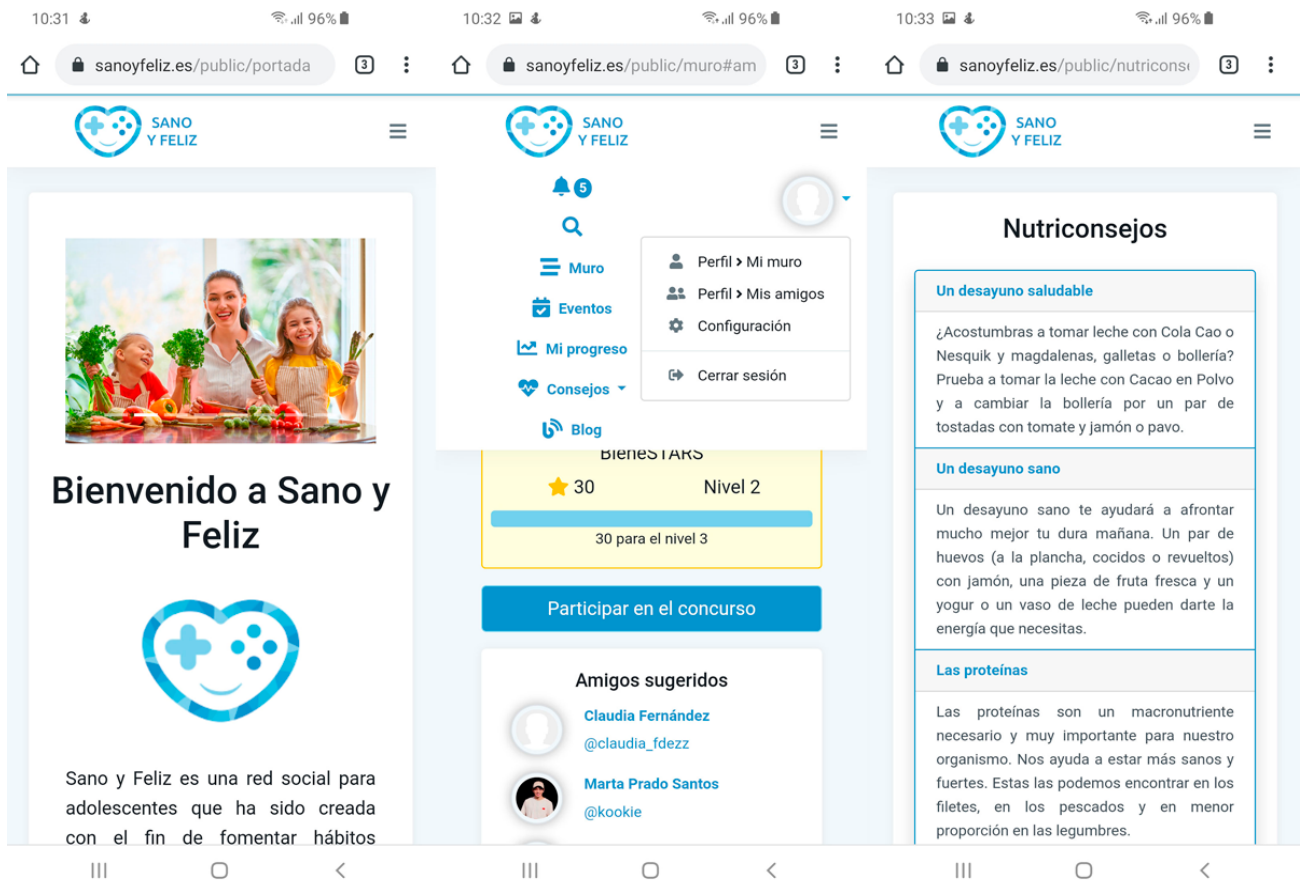

**Figure 2.** Screenshot of app showing healthyStars as a wall section (left panel) and in a user profile (right panel).

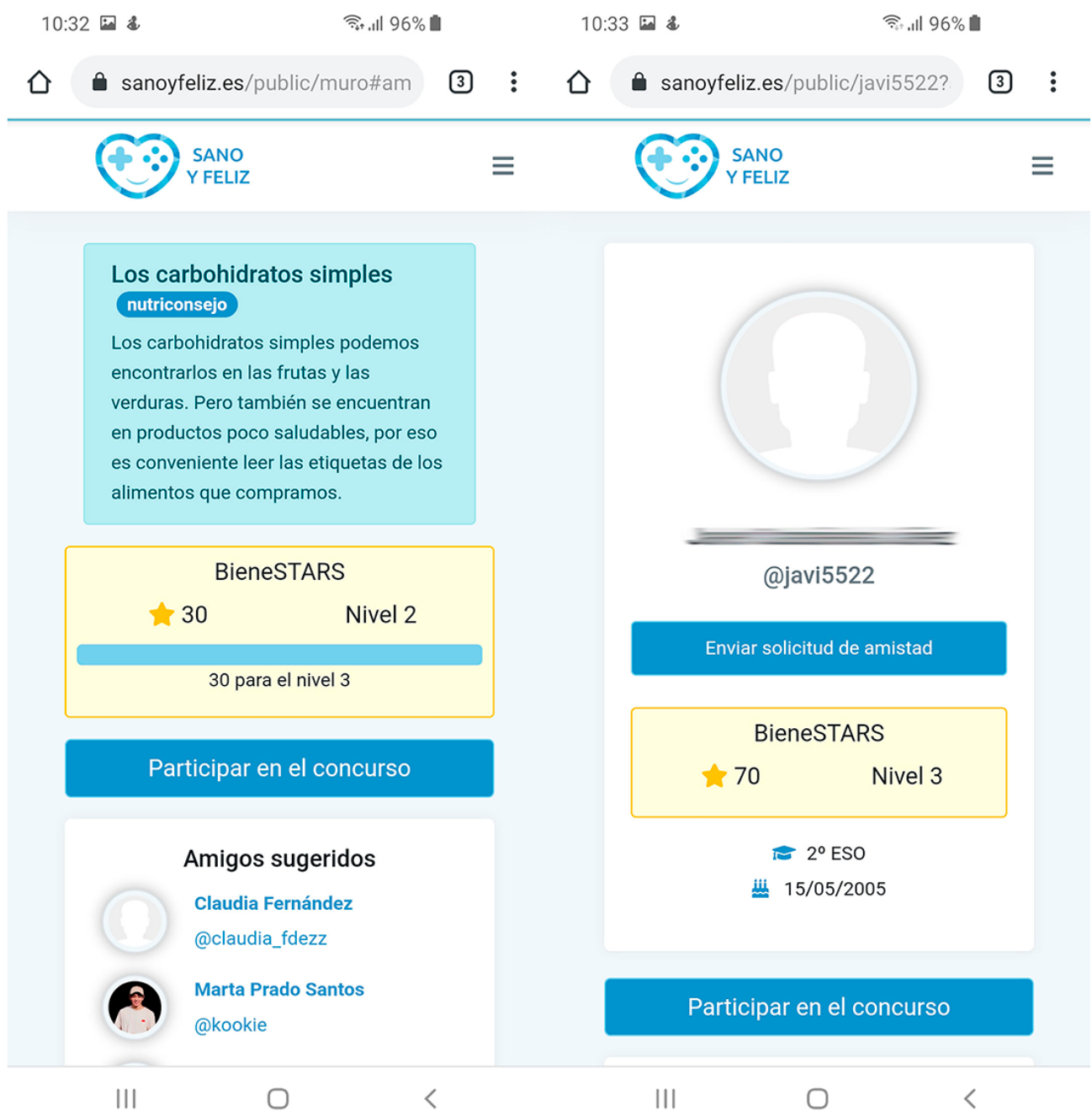

### Measuring Instruments and Variables

#### Anthropometric Variables

The anthropometric measurements collected at the beginning and end of the intervention were height and weight. Since the goal of the app was for adolescents to manage their own health, these data were self-referred, and parents could collaborate with their children when weighing and measuring themselves. A set of video tutorials was provided for clarification and to make the activity possible. With the height and weight data, the BMI was calculated, and the BMI percentile was determined according to age and gender, as per WHO guidelines [22]. In order to meet one of the proposed study objectives, a new dichotomous variable called the "percentile alteration" was created. This reflected whether individuals' BMIs had approached the 50th percentile (P50) at the end of the intervention. That is, if at the beginning of the intervention their BMI was at a percentile below or above the P50, and at the end of the intervention their BMI was at a percentile closer to the P50, they were considered to have improved their weight status.

The BMI percentile was chosen to compare the evolution of adolescents pre- and postintervention, as it is a widely used measure in studies of the pediatric population [23-25]. In addition, it is an objective measure that allows us to know whether adolescents have progressed in the intervention rather than to simply compare those who are underweight, overweight, and obese. To check whether the BMI percentile had improved after the intervention, a partition was applied to this variable, dividing it into 2, according to age and gender: (1) individuals at a BMI percentile below the P50, and (2) individuals at a BMI percentile above the P50. According to this partitioning, individuals with BMIs above the P50 in the initial stage would improve their weight status if they managed to reach a lower percentile, while individuals with BMIs below the P50 in the initial stage would improve their weight status if they managed to reach a higher percentile by the end of the intervention.

*Eating Habit Variable*

The questionnaire used to assess adherence to the Mediterranean diet was the Mediterranean Diet Quality Index (KIDMED) [26]. The KIDMED questionnaire has been the most widely used scoring system to estimate adherence to the Mediterranean diet in children and adolescents [27]. By means of 16 dichotomous questions, a total score of between 0 and 12 points can be obtained by which adherence to the Mediterranean diet can be classified, ranging from poor adherence to an optimal level of adherence (the higher the score, the better the quality of the diet). It was considered as a quantitative variable. The internal consistency of the scale in the study sample was high (Cronbach α=.71).

*Physical Activity Variable*

The level of physical activity was assessed by means of the Physical Activity Questionnaire for Adolescents (PAQ-A), which has been validated for the Spanish adolescent population between 12 and 17 years of age [28] and consists of 8 questions relating to the type and amount of physical exercise performed during the last 7 days. The total score ranges from 1 to 5 points, and the interpretation is that the higher the score, the more physical activity was performed. It was considered as a quantitative variable. The internal consistency of the scale in the study sample was high (Cronbach α=.76).

*Structural Variables*

Based on the methodology of the SNA, the "degree" (one of the measures that describe the centrality of the individual within the network) was studied and related to the variables corresponding to physical activity and diet and use of the app. "Degree" is defined as the number of reciprocal connections that the student has [29] (ie, I am your friend and you are my friend). In this study, the term "friendship" was applied when two people claimed to be friends with each other using the definition suggested by authors such as Damon et al [30]: "friendship is a reciprocal relationship that must be affirmed or recognized by both parties." The degree was obtained from both the initial network and the network generated at the end of the intervention, with the aim of studying the adolescents' social relations pre- and postintervention. In response to the statement, "From the following list, point out your closest friends," students were able to choose any classmates in their class with no limit to the quantity within the same course.

Furthermore, it was also necessary to identity the "leaders." Various studies on the application of the SNA paradigm to health interventions have attempted to identify these individuals, as they can act as facilitating agents for the dissemination of healthy behavior on the network following different strategies [31]. It was decided to use the degree as a centrality indicator. The dichotomous variable "opinion leader by degree" was then created. To identify the leaders, 15% of the individuals with higher degree values were chosen, following the recommendations of Valente [31], and individuals with the same score as the last of those included in the 15%, if any, were also included.

*Variables Related to Use of the eHealth App*

In order to study the activity of adolescents during their use of the eHealth app, a set of variables were used from the data collected from the interaction of the participants:

- number of entries;
- number of responses;
- number of likes;
- number of "healthyStars;" and
- number of interactions.

**Statistical Analysis**

The data were anonymized with the tool described by Benítez et al [32]. SPSS software (version 24.0; IBM Corp) was used for the statistical processing of the data obtained. For the analysis of descriptive data, frequencies and percentages were used for the qualitative variables, and the mean and standard deviation were used for the quantitative variables. A chi-square test was performed to verify whether there was a relationship between the groups, and the Student *t* test was used to compare mean scores between the groups. A repeated-measures analysis of variance (RM-ANOVA) was carried out to check the differences for time, group, and group-by-time interactions. UCINET software (version 6.679 [33]) was used for the calculation of the SNA measurements. The tests performed to study the normality of the distribution were the Kolmogorov-Smirnov test (for populations of more than 55 individuals) and the Shapiro-Wilk test (for populations of less than or equal to 55 individuals). The level of statistical significance was set at 0.05.

## *Results*

**Homogeneity of the CG and IG**

Before starting the statistical analysis, the homogeneity of both groups was checked in order to compare them and to ensure that differences observed in the study results between the groups were indeed the consequence of having used the app. An analysis of variance was performed in which age, BMI percentile according to age and gender, and both the KIDMED and PAQ-A scores were considered as dependent variables. Belonging to the CG or the IG was considered as an independent variable. As shown in Table 1, both groups were homogeneous. In the case of gender, the CG was composed of 52.7% (48/91) males versus 54.8% (115/210) males in the IG, with no significant differences ($P$=.75).

**Table 1.** Measures used to verify the homogeneity in the control group (CG) and the intervention group (IG).

| Measure | CG, mean (SD) | IG, mean (SD) | P value |
| --- | --- | --- | --- |
| Age | 12.77 (0.62) | 12.75 (0.72) | .76 |
| Age-adjusted BMI percentile | 57.13 (29.16) | 50.18 (30.18) | .07 |
| KIDMED[a] score | 7.31 (3.24) | 7.27 (2.48) | .92 |
| PAQ-A[b] score | 2.98 (0.90) | 2.77 (0.90) | .09 |

[a]KIDMED: Mediterranean Diet Quality Index.

[b]PAQ-A: Physical Activity Questionnaire for Adolescents.

## Difference Between the IG and the CG

An RM-ANOVA was performed to compare whether the pre-post intervention variation in the means of the four factors (changes in BMI percentile for individuals with BMIs below the P50 and those with BMIs above the P50, in PAQ-A score, and in KIDMED score) was significant, taking as a condition the group they belonged to (CG or IG).

Based on this analysis, it was determined that the changes in all of the study variables in both groups, along with the intervention time, were significantly different. The results obtained are shown in Table 2.

Only the effect of time on the complete study sample did not show significant pre-post intervention differences in changes in the adolescents with BMIs below the P50 ($F_1=0.041$, $P=.84$), but there were significant differences in the interaction of whether they belonged to the IG or the GC.

**Table 2.** Statistical changes in variables by time and by interaction between the control group (CG) and the intervention group (IG).

| Source | CG, mean (SD) | | IG, mean (SD) | | Sphericity assumed | $F_1$, time | P value, time | $F_1$, interaction | P value, interaction |
| --- | --- | --- | --- | --- | --- | --- | --- | --- | --- |
| | Pre[a] | Post[b] | Pre | Post | | | | | |
| **Initial age-adjusted BMI percentile** | | | | | | | | | |
| <P50[c] | 26.53 (12.61) | 24.19 (15.56) | 24.05 (13.92) | 26.25 (15.31) | 218.644 | 0.041 | .84 | 4.029 | .047 |
| >P50 | 78.09 (15.31) | 77.49 (15.87) | 77.31 (14.35) | 71.37 (18.77) | 506.456 | 16.488 | <.001 | 11.060 | .001 |
| PAQ-A[d] score | 2.98 (0.90) | 2.06 (1.53) | 2.77 (0.95) | 2.39 (1.45) | 8.776 | 60.291 | <.001 | 9.921 | .002 |
| KIDMED[e] score | 7.31 (3.24) | 7.47 (2.70) | 7.27 (2.48) | 8.11 (2.50) | 14.389 | 10.271 | .001 | 4.629 | .03 |

[a]Pre: preintervention.

[b]Post: postintervention.

[c]P50: 50th percentile.

[d]PAQ-A: Physical Activity Questionnaire for Adolescents.

[e]KIDMED: Mediterranean Diet Quality Index.

## Individual Study of the Results Obtained in the CG and IG

Using RM-ANOVA analysis, together with the study of the means obtained pre- and postintervention in each group, the improvement or worsening of each of the variables in each of the groups was analyzed.

In relation to variable 1 (age-adjusted BMI percentile for individuals with initial BMIs below the P50), there was a worsening in the CG, from a mean percentile of 26.53 (SD 12.61) to 24.19 (SD 15.56), while improvements were seen in the IG, from a mean percentile of 24.05 (SD 13.92) to 26.25 (SD 15.31); this difference between the groups was significant ($F_1=4.029$, $P=.047$). With respect to variable 2 (age-adjusted BMI percentile for individuals with initial BMIs above the P50), improvements were observed in the CG, from a mean percentile of 78.09 (SD 15.31) to 77.49 (SD 15.87), and also in the IG, from a mean percentile of 77.31 (SD 14.35) to 71.37 (SD 18.77), with a significantly greater improvement in the IG ($F_1=11.060$, $P=.001$).

After analyzing the PAQ-A scores, a worsening was observed in the CG, from 2.98 (SD 0.90) to 2.06 (SD 1.53), and also in the IG, from 2.77 (SD 0.95) to 2.39 (SD 1.45), but the worsening was significantly higher in the CG ($F_1=9.921$, $P=.002$).

Finally, the KIDMED score improved in the CG, from 7.31 (SD 3.24) to 7.47 (SD 2.70), and in the IG, from 7.27 (SD 2.48) to 8.11 (SD 2.50), with the improvement in the IG being significantly greater ($F_1=4.629$, $P=.03$).

In summary, the results showed significant improvements in the IG in all study variables, except PAQ-A score, where a worsening was detected, although it was significantly less than the worsening observed in the PAQ-A scores in the CG. It was also shown that the CG significantly worsened in the P50 approach variables in the age-adjusted BMI percentile for the group of individuals with BMIs below the P50 as well as in PAQ-A, while the P50 approach variables in the age-adjusted BMI percentile for the group of individuals with BMIs above the P50 and KIDMED scores improved, but to a significantly lesser extent than the improvement found in the IG.

### SNA Study

To carry out the SNA study, the four networks generated in the IG were analyzed:

- network 1: first-year students in the first school;
- network 2: second-year students in the first school;
- network 3: first-year students in the second school; and
- network 4: second-year students in the second school.

As can be seen in Table 3, the number of established reciprocal relationships (degree) increased significantly after use of the app. The distribution was not normal for any of the measures.

**Table 3.** Degree of the participants pre- and postintervention in the four networks generated in the intervention group.

| Network | Degree, mean (SD) | | P value |
|---|---|---|---|
| | Preintervention | Postintervention | |
| Network 1[a] | 7.66 (8.86) | 11.39 (8.74) | <.001 |
| Network 2[b] | 5.52 (3.90) | 7.64 (4.40) | <.001 |
| Network 3[c] | 4.55 (3.87) | 8.70 (4.54) | <.001 |
| Network 4[d] | 3.37 (2.89) | 5.68 (3.51) | <.001 |

[a]First-year students in the first school.

[b]Second-year students in the first school.

[c]First-year students in the second school.

[d]Second-year students in the second school.

Regarding the analysis of leaders by degree (15% of the sample with the highest values), the study of the distribution was not normal for any of the measures. The Wilcoxon test was used to analyze whether leaders were predominantly male or female and overweight or not, and whether the PAQ-A and KIDMED scores were higher in groups with leaders (leadership group [LG]) than in the rest of the sample (no leadership group [NLG]). No significance was found for gender (59.4% [19/32] male in the LG vs 53.9% [96/178] male in the NLG; $P$=.57), overweight (9.4% [3/32] in LG vs 11.2% [20/178] in NLG; $P$>.99), or KIDMED score ($P$=0.14). On the other hand, regarding the PAQ-A score, significant differences were obtained between the LG (3.33, SD 0.87) and the NLG (2.22, SD 1.47) ($P$=.002). In the analysis of app use by the leaders in comparison with the rest of the study sample, it was seen that the leaders used the app significantly more (Table 4).

**Table 4.** Analysis of the activity in "SanoYFeliz" of leaders in comparison with the other adolescents in the study sample.

| Activity | Leadership group, mean (SD) | No leadership group, mean (SD) | P value |
|---|---|---|---|
| Interactions | 12.66 (7.82) | 8.49 (10.86) | .045 |
| Entries | 4.25 (5.01) | 1.39 (3.56) | <.001 |
| Responses | 1.94 (3.82) | 0.54 (1.72) | <.001 |
| Likes | 3.44 (4.32) | 0.95 (3.02) | <.001 |
| HealthyStars | 47.66 (48.51) | 28.51 (34.06) | .008 |

## Discussion

### Principal Results

On analysis, it was found that use of the app helped users to achieve BMIs that were closer to the P50, both for the group of individuals with BMIs above the P50 and for those with BMIs below the P50. In this regard, many of the interventions for improving physical activity and eating habits in children and adolescents use the BMI value, age- and gender-adjusted BMI, z-score value, or age- and gender-adjusted percentile as a measure of effectiveness. In fact, these measures were the ones used to compare the effectiveness of interventions in two of the most comprehensive systematic reviews conducted to date on this type of intervention [29,34]. On the other hand, the explanation that we found for the greater significance of the group with BMIs above the P50 relates to the motivation of obese adolescents to live healthier lives. In fact, Silva et al [35] showed in their research that among the multiple motivational factors that an obese adolescent may have to lose weight, the desire for better health was especially important.

We were also able to see how the feeding in the IG improved significantly, something that did not happen in the CG. Similarly, the physical activity scores improved slightly in the IG but worsened significantly in the CG. These results indicate that the eHealth app, apart from being beneficial in helping adolescents reach an age-adjusted BMI percentile close to the average value, can also exert a positive influence on adolescents' behavior in relation to physical activity and diet.

In relation to the network leader and his or her relationship to healthy habits, this study found that being a leader and being physically active were significantly related. Since people with a high degree of centrality are a powerful channel of information, our study showed how leaders influenced their peers by increasing the physical activity levels of the rest of the students in the class. Our results are consistent with the literature, stating that friends have a great capacity to influence each other. Schofield et al [36] suggested that friendship, if reciprocal as was the case in our study, could exert a greater degree of influence when modifying a behavior related to physical activity. Similarly, Jago et al [37] found that in the case of boys, best friends influenced the physical activity that was performed, while in the case of girls, those who played sports with their best friends reported higher levels of activity. Macdonald-Wallis et al [38] went further and found not only that best friends and closest peers influenced adolescents' schoolwork, but also that a correlation exists between the behavior of young people with more friendship distance (ie, skipping a degree of friendship by corresponding with the friends of one's friends). In this regard, since our postintervention results reflect that the level of physical activity increased in the IG, we can affirm that these leaders have effectively influenced the rest of their peers in the performance of physical activity. Therefore, the intervention was effective not only in bringing the adolescents' BMI percentiles closer to P50 but also in modifying physical activity–related behavior, in turn weaving a support network among peers.

Also, the group leaders used the website app significantly more than the other adolescents in the study. In this sense, the variables "number of entries," "number of healthyStars," and "number of interactions" can be used as indicators of the level of participation and commitment to the use of the app, similar to what was observed by Tong et al [39]. Similarly, the "number of responses" and "number of likes" can be considered as a measure of the social support provided by each adolescent [40].

## Limitations

The research team is aware of the limitations of this study. One of the limitations is the fact that anthropometric measurements are self-referential. In this sense, although there are studies that reflect an underestimation of weight [41], there are other studies that defend that these data can be close to those measured by health professionals [42]. The lack of a longitudinal study to verify that these results can be perpetuated over time is also considered a limitation. Since this eHealth app is in an experimental phase, we can still improve it and measure it in later years if it proves to be as successful as expected. Another point to take into account is the cataloguing of leadership according to "degree," as the literature also recommends measuring "indegree" (nominations of friends received), "closeness" (closeness to other members of the network), and "betweenness" (capacity to mediate) to estimate leadership, without there being a clear consensus as to which of these measures is better for each specific case [43].

## Conclusions

In conclusion, it can be stated that the app was effective in helping its users bring their BMIs closer to the P50 for age and gender. Likewise, it is capable of modifying related behaviors or at least modelling them. On the other hand, the capacity of leaders to collaborate with these changes and to promote certain habits has been demonstrated. The eHealth app based on social networks can help in the fight against excess weight in the teenage population.

## Acknowledgments

This research was funded by the Junta de Castilla y León grant number LE014G18.

## Conflicts of Interest

None declared.

## Abbreviations

**CG:** control group
**IG:** intervention group
**KIDMED:** Mediterranean Diet Quality Index
**LG:** leadership group
**NLG:** no leadership group
**P50:** 50th percentile
**PAQ-A:** Physical Activity Questionnaire for Adolescents
**RM-ANOVA:** repeated-measures analysis of variance
**SNA:** social network analysis
**WHO:** World Health Organization